\documentclass{aastex631} 

\newcommand{\objname}{2015 VA$_{108}$}

\begin{document}

\title{New Active Asteroid 2015 VA108: A Citizen Science Discovery}




\correspondingauthor{Colin Orion Chandler}
\email{coc123@uw.edu}

\author[0000-0001-7335-1715]{Colin Orion Chandler}
\affiliation{Dept. of Astronomy \& the DiRAC Institute, University of Washington, 3910 15th Ave NE, Seattle, WA 98195, USA}
\affiliation{LSST Interdisciplinary Network for Collaboration and Computing, 933 N. Cherry Avenue, Tucson AZ 85721}
\affiliation{Dept. of Astronomy \& Planetary Science, Northern Arizona University, PO Box 6010, Flagstaff, AZ 86011, USA}

\author[0000-0001-5750-4953]{William J. Oldroyd}
\affiliation{Dept. of Astronomy \& Planetary Science, Northern Arizona University, PO Box 6010, Flagstaff, AZ 86011, USA}

\author[0000-0001-9859-0894]{Chadwick A. Trujillo}
\affiliation{Dept. of Astronomy \& Planetary Science, Northern Arizona University, PO Box 6010, Flagstaff, AZ 86011, USA}

\author[0000-0002-6023-7291]{William A. Burris}
\affiliation{Dept. of Physics, San Diego State University, 5500 Campanile Drive, San Diego, CA 92182, USA}
\affiliation{Dept. of Astronomy \& Planetary Science, Northern Arizona University, PO Box 6010, Flagstaff, AZ 86011, USA}

\author[0000-0001-7225-9271]{Henry H. Hsieh}
\affiliation{Planetary Science Institute, 1700 East Fort Lowell Rd., Suite 106, Tucson, AZ 85719, USA}
\affiliation{Institute of Astronomy and Astrophysics, Academia Sinica, P.O.\ Box 23-141, Taipei 10617, Taiwan}

\author[0000-0001-8531-038X]{Jay K. Kueny}
\altaffiliation{National Science Foundation Graduate Research Fellow}
\affiliation{University of Arizona Dept. of Astronomy and Steward Observatory, 933 North Cherry Avenue Rm. N204, Tucson, AZ 85721, USA}
\affiliation{Lowell Observatory, 1400 W Mars Hill Rd, Flagstaff, AZ 86001, USA}
\affiliation{Dept. of Astronomy \& Planetary Science, Northern Arizona University, PO Box 6010, Flagstaff, AZ 86011, USA}


\author[0000-0002-2204-6064]{Michele T. Mazzucato}
\altaffiliation{Active Asteroids Citizen Scientist}
\affiliation{Royal Astronomical Society, Burlington House, Piccadilly, London, W1J 0BQ, UK}

\author[0000-0002-9766-2400]{Milton K. D. Bosch}
\altaffiliation{Active Asteroids Citizen Scientist}

\author{Tiffany Shaw-Diaz}
\altaffiliation{Active Asteroids Citizen Scientist}

\begin{abstract}
We announce the discovery of activity, in the form of a distinct cometary tail, emerging from main-belt asteroid \objname{}. Activity was first identified by volunteers of the Citizen Science project Active Asteroids (a NASA Partner). We uncovered one additional image from the same observing run which also unambiguously shows \objname{} with a tail oriented between the anti-solar and anti-motion vectors that are often correlated with activity orientation on sky. Both publicly available archival images were originally acquired UT 2015 October 11 with the Dark Energy Camera (DECam) on the Blanco 4 m telescope at the Cerro Tololo Inter-American Observatory (Chile) 
as part of the Dark Energy Camera Legacy Survey. 
Activity occurred near perihelion and, combined with its residence in the main asteroid belt, \objname{} is a candidate main-belt comet, an active asteroid subset known for volatile sublimation.
\end{abstract}


\section{Introduction} \label{sec:intro}

Roughly 40 active asteroids have been discovered to date and, consequently, much about these objects remains unknown. Known for their unexpected display of cometary activity (i.e., tails, comae) despite being on asteroidal orbits (e.g., main-belt), active asteroids provide unique insights into dynamical, thermophysical, and astrochemical processes \citep{jewittActiveAsteroids2015}. Some bodies exhibit activity caused by volatile sublimation, such as the Main-belt comets (MBCs) found in the asteroid belt 
\citep{hsiehMainbeltCometsPanSTARRS12015}, and thus help us map the distribution of volatiles in the solar system. MBCs are also rare, with fewer than 15 identified thus far. With so few active asteroids and MBCs known, discovering additional active asteroids is crucial to gaining additional insights into these remarkable populations.

\clearpage
\section{Methods} \label{sec:methods}

\textit{Active Asteroids}\footnote{\url{http://activeasteroids.net}}, a NASA Partner program, is an online Citizen Science project (hosted on the Zooniverse\footnote{\url{https://www.zooniverse.org}} platform) that we created for the purpose of identifying more unusual active minor planets while engaging the public in the search \citep{chandlerChasingTailsActive2022}. Volunteers of the project look through images of small solar system bodies that we previously extracted from Dark Energy Camera (DECam) publicly available archival image data \citep{chandlerSAFARISearchingAsteroids2018,chandlerSixYearsSustained2019,chandlerCometaryActivityDiscovered2020b,chandlerRecurrentActivityActive2021,chandlerMigratoryOutburstingQuasiHilda2022}. We ask participants if they see activity or not, and we analyze their classification data to identify candidate objects. As of UT 2023 February 6, approximately 7,000 volunteers have carried out over 2.5 million classifications since the project launch in 2021 August, and our team has been consistently following up on promising candidates.

\section{Results} \label{sec:results}

\textit{Active Asteroids} volunteers identified an image of \objname{} (semi-major axis $a=3.13$~au, eccentricity $e=0.22$, inclination $i=8.50^\circ$, perihelion distance $q=2.45$~au, aphelion distance $Q=3.81$~au, Tisserand parameter with respect to Jupiter $T_\mathrm{J}=3.160$; retrieved UT 2023 January 27 from JPL Horizons; \citealt{giorginiJPLOnLineSolar1996}) as showing activity. Our team carried out additional searches through archival image data and found one additional image of \objname{} acquired during the same observing run as the image that prompted our study.

\begin{figure}[h]
    \centering
    \begin{tabular}{cc}
        \includegraphics[width=0.47\linewidth]{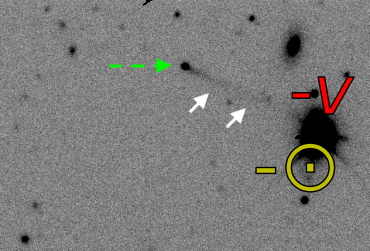} & \includegraphics[width=0.47\linewidth]{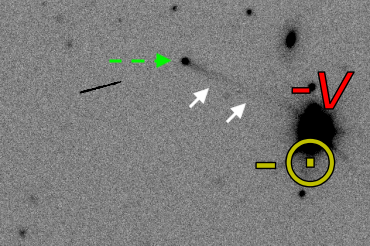} \\
    \end{tabular}
    \caption{\objname{} (dashed arrow) with a pronounced tail (white arrows) oriented between the anti-motion ($-v$) and anti-solar ($-\odot$) directions. 
    These 114~s $r$-band (left) and 125~s $g$-band (right) DECam images were captured with the Blanco 4~m telescope (Cerro Tololo Inter-American Observatory, Chile) on UT 2015 October 11 (Program 2014B-0404, PIs Schlegel and Dey, observers D. James, A. Dey, A. Patej). 
    }
    \label{fig:activity}
\end{figure}

Figure \ref{fig:activity} shows \objname{} with a conspicuous tail approximately oriented between the anti-solar and anti-motion vectors as projected on the plane of the sky. At the time (UT 2015 October 11), \objname{} was at a heliocentric distance $r_h=2.44$~au, outbound from its recent perihelion passage. \objname{} qualifies as an MBC candidate because (1) activity occurred near perihelion, and (2) \objname{} has an orbit bound to the main asteroid belt.

\clearpage

\section*{Acknowledgements}
\begin{acknowledgments}


\textbf{General:} Many thanks to Dr.\ Mark Jesus Mendoza Magbanua (University of California San Francisco) for continued feedback about the project and its findings. 

\textbf{Citizen Science:} We thank Elizabeth Baeten (Belgium) for moderating the Active Asteroids forums. We thank our NASA Citizen Scientists that examined \objname{}: 
Al Lamperti	(Royersford, USA), 
Melany Van Every (Lisbon, USA), 
Michele T. Mazzucato (Florence, Italy), 
Milton K. D. Bosch, MD (Napa, USA), 
Nazir Ahmad	(Birmingham, UK), 
Steven Green (Witham, UK), 
Tiffany Shaw-Diaz (Dayton, USA), 
 and 
Timothy Scott (Baddeck, Canada). 
%
We also thank super-classifier %
Marvin W. Huddleston (Mesquite, USA)
. 
Many thanks to Cliff Johnson (Zooniverse) and Marc Kuchner (NASA) for their ongoing guidance.

\textbf{Funding:} This material is based upon work supported by the NSF Graduate Research Fellowship Program under grant No.\ 2018258765 and grant No.\ 2020303693. 
C.O.C., H.H.H., and C.A.T.\ acknowledge support from the NASA Solar System Observations program (grant 80NSSC19K0869). W.J.O. acknowledges support from NASA grant 80NSSC21K0114. 
This work was supported in part by NSF awards 1950901 (NAU REU program in astronomy and planetary science). 
Computational analyses were run on Northern Arizona University's Monsoon computing cluster, funded by Arizona's Technology and Research Initiative Fund.

\textbf{Software \& Services:} 
World Coordinate System corrections facilitated by \textit{Astrometry.net} \citep{langAstrometryNetBlind2010}. 
This research has made use of 
NASA's Astrophysics Data System, 
the Institut de M\'ecanique C\'eleste et de Calcul des \'Eph\'em\'erides SkyBoT Virtual Observatory tool \citep{berthierSkyBoTNewVO2006}, 
and 
data and/or services provided by the International Astronomical Union's Minor Planet Center, 
SAOImageDS9, developed by Smithsonian Astrophysical Observatory \citep{joyeNewFeaturesSAOImage2006}. 

\textbf{Facilities \& Instrumentation:} This project used data obtained with the Dark Energy Camera (DECam), which was constructed by the Dark Energy Survey (DES) collaboration. 
This research uses services or data provided by the Astro Data Archive at NSF's NOIRLab. 
Based on observations at Cerro Tololo Inter-American Observatory, NSF’s NOIRLab (NOIRLab Prop. ID 2014B-0404; PI: D. Schlegel). 
%
%
The Legacy Surveys consist of three individual and complementary projects: the DECam Legacy Survey (DECaLS; Proposal ID \#2014B-0404; PIs: David Schlegel and Arjun Dey), the Beijing-Arizona Sky Survey (BASS; NOAO Prop. ID \#2015A-0801; PIs: Zhou Xu and Xiaohui Fan), and the Mayall z-band Legacy Survey (MzLS; Prop. ID \#2016A-0453; PI: Arjun Dey). 

\end{acknowledgments}

\vspace{5mm}
\facilities{
CTIO:4m (DECam) 
}


\software{
        astropy \citep{robitailleAstropyCommunityPython2013}, 
        {\tt Matplotlib} \citep{hunterMatplotlib2DGraphics2007},
        {\tt NumPy} \citep{harrisArrayProgrammingNumPy2020},
        {\tt pandas} \citep{rebackPandasdevPandasPandas2022}, 
        {\tt SAOImageDS9} \citep{joyeNewFeaturesSAOImage2006},
        {\tt SciPy} \citep{virtanenSciPyFundamentalAlgorithms2020}
          }

\clearpage
\bibliography{zotero}{}
\bibliographystyle{aasjournal}



\end{document}